\documentclass[11pt]{article}

\usepackage{dialogue}

\usepackage{ifxetex}
\ifxetex
    \usepackage{fontspec}
    \setromanfont{Times New Roman}
\else
    \usepackage[T2A]{fontenc}
    \usepackage[utf8]{inputenc} 
    \usepackage[british, russian]{babel}
    \usepackage{newtxtext,newtxmath}
\fi

\usepackage[a4paper,top=2.5cm,bottom=2.5cm,left=2.5cm,right=2.5cm]{geometry}
\usepackage{url}
\usepackage{adjustbox}
\usepackage{pgf}
\usepackage{float}
\usepackage{booktabs}
\usepackage{microtype}
\usepackage{amsmath}
\usepackage{pifont}

\usepackage{array}
\graphicspath{{media/}} 
\usepackage{biblatex}
\usepackage{multicol, multirow}
\usepackage{hyperref}

\dialogfinalcopy 

\usepackage{graphicx}

\title{Building Russian Benchmark for  Evaluation \\of Information Retrieval Models}

\begin{document}
\selectlanguage{british}

\maketitle
\vspace{-1mm}
\begin{center}
    \begin{minipage}[t]{0.45\linewidth}
    \centering
    \textbf{Grigory Kovalev}\\
    Lomonosov Moscow\\
    State University\\
    Russia\\
    \texttt{kaengreg@ya.ru}
    \end{minipage}
    \hfill
    \begin{minipage}[t]{0.45\linewidth}
    \centering
    \textbf{Mikhail Tikhomirov}\\
    Lomonosov Moscow\\
    State University\\
    Russia\\
    \texttt{tikhomirov.mm@gmail.com}
    \end{minipage}

    \vspace{1em}
    \begin{minipage}[t]{0.45\linewidth}
    \centering
    \textbf{Evgeny Kozhevnikov}\\
    Lomonosov Moscow\\
    State University\\
    Russia\\
    \texttt{dovvakkin@gmail.com}
    \end{minipage}
    \hfill
    \begin{minipage}[t]{0.45\linewidth}
    \centering
    \textbf{Max Kornilov}\\
    Lomonosov Moscow\\
    State University\\
    Russia\\
    \texttt{max.korn@bk.ru}
    \end{minipage}

    \vspace{1em}
    \begin{minipage}[t]{0.45\linewidth}
    \centering
    \textbf{Natalia Loukachevitch}\\
    Lomonosov Moscow\\
    State University\\
    Russia\\
    \texttt{louk\_nat@mail.ru}
    \end{minipage}
\end{center}
\vspace{1mm}

\begin{abstract}
  We introduce RusBEIR, a comprehensive benchmark  designed for zero-shot evaluation of information retrieval (IR) models in the Russian language. Comprising 17 datasets from various domains, it integrates adapted, translated, and newly created datasets, enabling systematic comparison of lexical and neural models.  Our study highlights the importance of preprocessing for lexical models in morphologically rich languages and confirms BM25 as a strong baseline for full-document retrieval. Neural models, such as mE5-large and BGE-M3, demonstrate superior performance on most datasets, but face challenges with long-document retrieval due to input size constraints.  RusBEIR offers a unified, open-source framework that promotes research  in Russian-language information retrieval.   
 The benchmark is available for public use on \href{https://github.com/kaengreg/rusBeIR}{GitHub}.
 
  \textbf{Keywords}: information retrieval, benchmark, lexical model, dense model, reranker 
\end{abstract}

\selectlanguage{russian}

\begin{center}
\russiantitle{Создание русского бенчмарка для  оценки моделей \\информационного поиска}
  \medskip \setlength\tabcolsep{2cm}
  \begin{tabular}{cc}
    \textbf{Ковалев Григорий} & \textbf{Тихомиров Михаил} \\
    МГУ им. М.В. Ломоносова & МГУ им. М.В. Ломоносова\\
    Россия  & Россия \\
    {\tt kaengreg@ya.ru} &  {\tt tikhomirov.mm@gmail.com} \\
  \end{tabular}
  \medskip

\medskip \setlength\tabcolsep{2cm}

\begin{tabular}{cc}
    \textbf{Кожевников Евгений} & \textbf{Корнилов Максим} \\
     МГУ им. М.В. Ломоносова & МГУ им. М.В. Ломоносова\\
     Россия  & Россия \\
     {\tt dovvakkin@gmail.com} & {\tt max.korn@bk.ru}\\
    \end{tabular}
\medskip

\begin{tabular}{c}
    \textbf{Лукашевич Наталья} \\
     МГУ им. М.В. Ломоносова\\
     Россия \\
     {\tt louk\_nat@mail.ru}\\
    \end{tabular}
\medskip

\end{center}

\begin{abstract}
  Мы представляем RusBEIR — это бенчмарк, предназначенный для zero-shot оценки моделей информационного поиска (IR) на русском языке. Он включает 17 наборов данных из различных доменов, объединяя адаптированные, переведенные и  созданные наборы данных, что позволяет проводить систематическое сравнение лексических и нейронных моделей.   Наше исследование подчеркивает важность предобработки для лексических моделей в языках с богатой морфологией и подтверждает, что модель BM25  обеспечивает высокое качество поиска, особенно для полных документов. Нейронные модели, такие как mE5-large и BGE-M3, показывают высокие результаты на большинстве наборов данных, но сталкиваются с трудностями при работе с длинными документами из-за ограничений на максимальную длину входа.   RusBEIR предоставляет унифицированную открытую платформу, способствующую развитию исследований в области информационного поиска на русском языке.   RusBEIR является проектом с открытым исходным кодом и доступен на \href{https://github.com/kaengreg/rusBeIR}{GitHub}.
  
  \textbf{Ключевые слова:} информационный поиск, бенчмарк, лексическая модель, нейросетевая модель, реранкер
\end{abstract}
\selectlanguage{british}

\section{Introduction}
Traditionally, Information Retrieval (IR) was based on lexical models such as TF-IDF and BM25, but these models are known as bag-of-words models, which do not take into account the document context. Modern approaches are based on neural models, in particular on Transformer models. Most recent advancements in IR leverage neural retrieval models built upon pre-trained Transformer architectures, such as BERT \cite{devlin-etal-2019-bert}. These models address the limitations of traditional lexical methods by capturing semantic relationships and contextual information, enabling them to bridge the lexical gap inherent in keyword-based retrieval approaches. Unlike lexical models, which rely solely on the presence of query terms in documents, neural models represent queries and documents in a dense vector space, facilitating more accurate retrieval through similarity measures like cosine similarity.

Neural retrieval systems have demonstrated significant performance improvements over traditional methods, particularly in tasks such as open-domain question answering, claim verification, and passage retrieval. However, these advancements often come at the cost of increased computational resources and the need for extensive training data. Due to the scarcity of labeled data, neural models are frequently applied in zero-shot settings.

Traditionally, neural retrievers have been trained on large datasets such as MS MARCO \cite{bajaj2016ms} and Natural Questions \cite{hardeniya2016natural}. Before the introduction of BEIR \cite{thakur2021beir}, these models were often evaluated on the same datasets they were trained on, gaining a significant advantage over lexical approaches like BM25. 

To address this limitation, the authors of BEIR introduced a robust and diverse benchmark designed to evaluate model generalization across tasks and domains. BEIR consists of 18 retrieval datasets from a variety of domains, providing a more accurate and comprehensive framework for evaluating neural retrieval systems. Notably, the results of BEIR revealed that neural models do not consistently outperform lexical approaches, highlighting the need for careful evaluation in diverse settings.

The zero-shot application of neural retrievers is particularly important for underrepresented languages, such as Slavic languages, where the availability of information retrieval datasets is limited. Consequently, there is a growing demand for multilingual evaluation benchmarks akin to the monolingual BEIR framework. Such benchmarks would enable robust cross-lingual evaluation and foster the development of neural retrieval systems for less commonly studied languages, addressing the current gaps in multilingual information retrieval research.

Interestingly, the performance gap between lexical and dense retrieval models remains a topic of interest. Although dense models typically excel in retrieval accuracy, lexical methods such as BM25 offer a lightweight alternative with significantly lower computational overhead. Investigating this trade-off can provide valuable insight into the practical application of retrieval models across diverse scenarios, especially when computational efficiency is a priority.

In this paper, we present RusBEIR, a BEIR-inspired benchmark designed for the zero-shot evaluation of Information Retrieval (IR) models in the Russian language. RusBEIR comprises 17 datasets that span various domains and tasks. Some datasets have been adapted from BEIR and similar benchmarks, others are newly collected specifically for this benchmark or sourced from existing Russian or multilingual benchmarks.
Our primary objective is to establish a large-scale benchmark tailored to information retrieval in Russian, with a particular emphasis on zero-shot approaches. In addition, we explore whether neural models consistently outperform traditional lexical methods across diverse scenarios. With this aim, we evaluate a range of models, including BM25, BGE-M3, mE5, RoSBERTa, and LaBSE \cite{chen2024bge,wang2024multilingual,snegirev2024russian,feng2022language}.
This benchmark offers a comprehensive resource for advancing and evaluating IR systems in Russian, fostering research into the comparative strengths of neural and lexical approaches.

 \section {Related work}

 The information-retrieval domain has a rich history of creating datasets, benchmarks, organizing various evaluations. Specialized evaluation conferences such as TREC, CLEF, NTCIR have been held since 90-th of 20 century.  There exist numerous national information-retrieval initiatives: in Poland \cite{kobylinski2023poleval}, India \cite{ganguly2023proceedings} and other countries. In Russia during 2003-2011, ROMIP workshop \cite{dobrov2004russian} was a place for evaluation of approaches in information-retrieval tasks, such as ad hoc retrieval, thematic categorization, question-answering, summarization, etc.

 Evaluating neural models in information retrieval requires creating benchmarks comprising diverse datasets. Development of information-retrieval benchmarks for non-English languages is usually based on the English BEIR benchmark \cite{thakur2021beir}. 
 
 The Polish benchmark BEIR-PL \cite{wojtasik2024beir} was created via automatic translation of  13 datasets from  BEIR. For translation,  the Google Translate service was used. In \cite{dadas2024pirb}, the authors describe the Polish Information Retrieval Benchmark (PIRB),  encompassing 41 text information retrieval tasks for Polish.  The datasets in PIRB comprise the BEIR-PL datasets, several other existing information-retrieval datasets, and also nine datasets crawled from Polish websites. In evaluation, it was found that the best results on the benchmark were achieved by the mE5-large model \cite{wang2024multilingual}. The authors also trained a learning-to-rank model combining scores of several basic models and achieved better results.

 To create Dutch BEIR, the authors of \cite{banar2024beir} translated initial BEIR datasets into Dutch using the Gemini-1.5-flash model. To assess the translation quality,  ten items from each dataset were randomly sampled and checked by a Dutch native speaker. It was shown that  98\% of checked samples were translated correctly or with minor issues. The authors tested BM25, neural models (including  mE5-large and BGE-M3 models) and reranking approaches combining BM25 and a neural reranker. They conclude that BM25 still provides a competitive baseline, and, in many cases, is only outperformed by larger dense models.

 The authors of \cite{acharya2024hindi} created Hindi BEIR benchmark. Hindi-BEIR encompasses 15 diverse datasets from 6 distinct domains. They translated BEIR datasets using  Indic-Trans2 model, a neural translation model supporting translations across all 22  Indic languages (including English). They translated 9 datasets from the source BEIR benchmark. To check the quality of translation, the authors  back-translated the Hindi translations into English. Then they calculated the char-based Chrf(++) score \cite{popovic2015chrf} between the original English query/document and the backtranslated English query/document. Also 5 publicly available information-retrieval datasets were added to the benchmark. In experiments, neural models (BGE-M3, mE5, LASER, LaBSE) were compared with BM25. The best results were obtained with BGE-M3, which is signifcantly better than other approaches.

 For Russian,  the MTEB benchmark \cite{muennighoff2023mteb} for evaluating embeddings has been created. Russian MTEB comprises 23 datasets in 7 task categories including three information-retrieval datasets \cite{snegirev2024russian}.  The best models in the Russian MTEB information-retrieval section with the size less than 1b are BGE-M3 \cite{chen2024bge} and Multilingual E5-large \cite{wang2024multilingual}.

 \section{Datasets in RusBEIR}
 RusBEIR is a Russian benchmark inspired by BEIR \cite{thakur2021beir}, designed for zero-shot evaluation of Information Retrieval (IR) models. Adhering to the principles of BEIR, it offers a robust and diverse evaluation framework, enabling the assessment of IR models across a wide range of tasks and domains in the Russian language.

 The datasets in the RusBEIR benchmark consist of available open-source datasets, datasets that have been translated from English, and newly created datasets. Table \ref{tab:datasets} provides a description of the available datasets. We will discuss the datasets in more detail in the following subsections.


 \subsection{Translated BEIR Datasets}
    BEIR consists of multilingual and monolingual (English) datasets. To achieve reproducibility of results from BEIR and its analogues, it was decided to translate the monolingual datasets into the Russian language and evaluate them with models used in our benchmark.
    
   The choice of translation method was based on studies conducted as part of the creation of the multilingual MsMarco dataset mMarco \cite{bonifacio2021mmarco}, where experiments with Google Translate and the Helsinki model were conducted, and the results of similar experiments from the PL-BEIR \cite{wojtasik2024beir} project were analyzed.
    According to the results of these studies, Google Translate showed better translation quality compared to the Helsinki model. Therefore, Google Translate was chosen.
    
    
    

    As a result, we introduce 4 datasets from the original BEIR datasets \cite{thakur2021beir}, which were translated into the Russian language.

    \begin{itemize}
        \item 
         \textbf{NF-Corpus} is a comprehensive full-text English retrieval dataset designed for medical information retrieval tasks. It contains a collection of queries formulated in non-technical English sourced from NutritionFacts.org \footnote{NutritionFacts.org} and corresponding medical documents written in a complex terminology-heavy language primarily derived from PubMed \footnote{https://pubmed.ncbi.nlm.nih.gov}, a database of medical literature. 
        \item \textbf{ArguAna} is a dataset designed for the argument retrieval task, derived from debates on idebate.org \footnote{idebate.org}. It covers controversial topics across 15 themes, such as “economy” and “health.” The dataset includes a corpus consisting of debate texts and queries derived from these debates. The task is to retrieve relevant arguments from the corpus.
        \item \textbf{SciFact} is a dataset for scientific claim verification, consisting of expert-written claims paired with abstracts from research literature. Each abstract is annotated with evidence supporting or refuting the claims, along with rationales justifying the decisions.
        \item \textbf{SCIDOCS} is a dataset focused on citation prediction, designed to evaluate the ability of scientific document embeddings to predict citation relationships between research papers.
    \end{itemize}

 \begin{table}[h!]
\centering
\renewcommand{\arraystretch}{1.25}
\resizebox{\linewidth}{!}{
\begin{tabular}{@{}llcc|cccccc@{}}
\toprule
\textbf{Source (↓)} & \textbf{Task (↓)} & \textbf{Dataset (↓)} & \textbf{Origin (↓)} & \textbf{Relevancy} & \textbf{Train} & \textbf{Dev} & \textbf{Test} & \textbf{Corpus} & \textbf{Avg. Word Lengths (D/Q)} \\ 
\hline
BEIR & Bio-Medical IR  & rus-NFCorpus & Translation & Binary & 2,590 & 324 & 323  & 3,633 & 216.6 / 3.5 \\ 
BEIR & Argument Retrieval & rus-ArguAna & Translation & Binary & --- & --- & 1,406 & 8,674 &  147.8 / 173.8 \\
BEIR & Fact Checking&  rus-SciFact & Translation & Binary & 809 & --- & 300 & 5,183 & 185.8 / 11.2 \\
BEIR & Citation-Prediction &  rus-SCIDOCS & Translation & Binary & --- & --- & 1000 & 25,657 & 153.1 / 9.8 \\
\hline
BEIR & Information-Retrieval  & rus-MMARCO  & Part of multilingual & Binary & 502,939 & 6980 & --- & 8,841,823 & 49.6 / 5.95\\ 
Open-Source Dataset & Information-Retrieval  & rus-MIRACL  & Part of multilingual & Binary &  4,683 & 1,252 & --- & 9,543,918 & 43 / 6.2 \\ 
Open-Source Dataset & Question Answering (QA)  & rus-XQuAD &  Part of multilingual & Binary & --- & 1,190 & --- & 240 &  112.9 / 8.6  \\ 
Open-Source Dataset & Question Answering (QA)  & rus-XQuAD-sentences &  Part of multilingual & Binary & --- & 1,190 & --- & 1212 &  22.4 / 8.6 \\
Open-Source Dataset & Question Answering (QA)  & rus-Tydi QA &  Part of multilingual & Binary & --- & 1,162 & --- & 89,154 & 69.4 / 6.5  \\ 
\hline
Open-Source Dataset & Information-Retrieval  & SberQuAD-retrieval & Originally Russian & Binary &  45,328 & 5,036 & 23,936 & 17,474 & 100.4 / 8.7  \\ 
Open-Source Dataset & Information-Retrieval  & ruSciBench-retrieval & Originally Russian & Binary &  --- & 345 & --- & 200,532  & 89.9 / 9.2 \\ 
Open-Source Dataset & Question Answering (QA) & ru-facts &  Originally Russian & Binary & 2,241 & 753 & --- & 6,236 & 28.1 / 23.9  \\
RU-MTEB & Information-Retrieval &  RuBQ  & Originally Russian & Binary &  --- & --- & 1,692 & 56,826 & 62.07 / 6.4 \\ 
RU-MTEB & Information-Retrieval &  Ria-News & Originally Russian & Binary &  --- & --- & 10,000 & 704,344 & 155.2 / 8.8 \\ 
\hline
rusBEIR & Information-Retrieval & wikifacts-articles & Originally Russian & 3-level &  --- & 540 & --- & 1,324 & 2,535.9 / 11.4 \\ 
rusBEIR & Fact Checking &  wikifacts-para & Originally Russian & 3-level &  --- & 540 & --- &  15,317 & 219.2 / 11.4 \\ 
rusBEIR & Information-Retrieval  & wikifacts-sents & Originally Russian & 3-level &  --- & 540 & --- & 188,026 & 17.8 / 11.4 \\ 
\bottomrule
\end{tabular}
}
\caption{Overview of datasets and tasks for information retrieval and related fields. All datasets are available at \href{https://huggingface.co/collections/msu-rcc-lair/rusbeir-datasets-6720fb076978ab6a77f4f64c}{HuggingFace}}
\label{tab:datasets}
\end{table}

 \subsection{Russian Parts of Multilingual Datasets}
  The main objective of BEIR is to gather a large and diverse set of data from various domains and tasks. This will force models to operate in an out-of-distribution environment and help to evaluate them more accurately. In order to expand our collection of Russian datasets, we also retrieved the Russian portions of existing multilingual datasets, including mMARCO \cite{bonifacio2021mmarco}, MIRACL \cite{zhang2023miracl}, XQUAD \cite{artetxe2020cross}, and TyDiQA \cite{clark2020tydi}.

  The \textbf{mMARCO} (Multilingual MS MARCO) dataset \cite{bonifacio2021mmarco} is a multilingual adaptation of the popular MS MARCO dataset, designed for information retrieval and question answering tasks. It extends the original English MS MARCO dataset into multiple languages, including Russian.
    
  \textbf{MIRACL} is a multilingual dataset for information
 retrieval in 18 languages. The queries were taken mainly from the Mr. TYDI dataset. Passages were retrieved from Wikipedia by an ensemble model, and 10 top documents were annotated by human annotators.

  \textbf{XQuAD} (Cross-lingual Question Answering Dataset) is a benchmark dataset designed to evaluate the performance of cross-lingual question answering systems. It consists of a collection of 240 passages and 1,190 question-answer pairs from the development set of the SQuAD v1.1 dataset \cite{rajpurkar2016squad}, along with their professional translations into 10 languages: Spanish, German, Greek, Russian, Turkish, Arabic, Vietnamese, Thai, Chinese, and Hindi. This makes the dataset entirely parallel across 11 languages.

  \textbf{Tydi QA} is a question-answering dataset covering 11 typologically
  diverse languages. Questions were written by humans on Wikipedia topics. Answers should not be contained  in the first 100 characters of the corresponding Wikipedia article. The questions were written for each language, not translated.

 \subsection{Existing Russian Datasets}
 The Russian Massive Text Embedding Benchmark (ruMTEB) is an extension of the Massive Text Embedding Benchmark (MTEB) tailored specifically for the Russian language. The authors of ruMTEB introduced 17 new datasets in Russian which were categorized into 7 groups. 
 
 In our benchmark we use 2 of presented IR datasets: RuBQ and Ria-News.  
 \textbf{RuBQ} \cite{rybin2021rubq} is a specialized dataset for Russian-language question answering over Wikidata, offering a rich set of questions paired with structured answers. 
 
 The \textbf{Ria-News} dataset \cite{gavrilov2019self} is a collection of Russian-language news articles published by the RIA Novosti news agency (2010-2014). This dataset presents a task in which a model is required to locate the text of a specific news article within a larger corpus of news articles based on its corresponding title, which acts as a query.

  Besides, we added publicly-available IR-related datasets: SberQuad \cite{efimov2020sberquad}, ruSciBench and ru-facts \cite{kozlova2023fact}.  
  
  \textbf{SberQuAD} is a Russian-language machine reading comprehension (MRC) dataset inspired by the popular English SQuAD \cite{rajpurkar2016squad} (Stanford Question Answering Dataset).  It provides annotated passages and question-answer pairs in Russian.
  
  \textbf{ruSciBench} is a Russian-language benchmark designed to evaluate the performance of text embedding models for scientific articles.
 \footnote {The dataset is a ported version of qa\_science\_ru {https://huggingface.co/datasets/AIR-  Bench/qa\_science\underline{}\_ru} from the Air-Bench repository, which in turn is a port of the ru\_sci\_bench (https://huggingface.co/datasets/mlsa-iai-msu-lab/ru\_sci\_bench) dataset from the MSLA-Iai MSU Lab} The corpus consists of abstracts, and the queries are LLM-generated questions for these abstracts.

 
 \textbf{ru-facts} \cite{kozlova2023fact} is a fact-checking dataset developed by translating and expanding the FEVER dataset with additional data from the Russian news summarization corpus Gazeta \footnote{https://huggingface.co/datasets/IlyaGusev/gazeta}, using a paraphrasing model \footnote{https://habr.com/en/company/sberdevices/blog/667106/}, and rule-based transformations from the Ru\_Paraphraser dataset \footnote{https://huggingface.co/datasets/merionum/ru\_paraphraser}.

 \subsection{New Russian Wikipedia-based Datasets}

 We also introduce a new series of Russian Wikipedia-based datasets. The  datasets are based on Wikipedia section “Did you know ...”. The section contains interesting facts, which are extracted from Wikipedia articles. The articles mentioned in a fact are provided with hyperlinks. 
For example, the fact “The 2024 American Samoan gubernatorial election was won by Pula and Pulu?” mentions three Wikipedia articles ("2024 American Samoan gubernatorial election", "Pula", "Pulu"),   from which the fact should be inferred. 

University students were asked to find relevant sentences in the mentioned articles that confirm the fact. 
They marked relevant sentences with scores of 2 or 1. Irrelevant sentences have zero scores. Relevant sentences with score 2 contain the full fact. If a sentence contains a part of the fact it obtains score 1. In total, 540 facts have been annotated.

 Using facts, extracted articles and created annotations, three datasets with the same queries but different documents have been created.
 \begin{itemize}
     \item  \textbf{wikifacts-sents} dataset consists of sentences extracted from articles, some of which confirm the fact which stands as a query. The documents in this dataset are the shortest in the benchmark;
     \item \textbf{wikifacts-articles} dataset comprises all full articles mentioned in facts. Relevant articles contain relevant sentences. This dataset includes the longest documents in the benchmark and can be used for evaluation of full-document retrieval;
     \item \textbf{wikifacts-para} dataset comprises existing paragraphs from the extracted articles, the documents on the datasets are significantly shorter than in the  wikifacts-articles dataset, but still longer than most benchmark datasets;

 \end{itemize}

Having such variants, we can evaluate different information-retrieval tasks on the same annotated data.

\subsection{BEIR Format Compatibility}
  Our datasets are presented in a unified format and are compatible with the original BEIR benchmark.  Queries are predetermined questions in natural language that are used to evaluate the performance of information retrieval (IR) systems. A corpus refers to a collection of documents that the system searches through in order to find relevant information for the given query. Relevance judgments, also known as qrels, indicate the association between queries and documents. All queries, corpora, and relevance judgments are stored in JSONL and TSV file formats, respectively.

 \section{Models}
 For evaluation, we used the BM25 lexical model and dense retrieval models. 

 \subsection{Preprocessing for BM25 model}
 The main baseline was calculated using the BM25 lexical model implemented in the Elasticsearch engine\footnote{https://www.elastic.co/}, with the language analyzer disabled to avoid stemming, which is less suitable for the Russian language. We specially preprocessed data to be used as input for BM25.

 The text preprocessing method consists of the following steps:
    \begin{enumerate}
        \item \textbf{Lowercasing}: Converting all text to lowercase to ensure uniformity and eliminate case sensitivity.
        \item \textbf{Punctuation and Special Character Removal}: Using regex to remove non-alphanumeric characters, leaving only letters, digits, and spaces to reduce noise.
        \item \textbf{Space Normalization}: Removing extra spaces and trimming leading or trailing whitespace.
        \item \textbf{Tokenization}: Splitting text into individual words for processing.
        \item \textbf{Lemmatization}: Using PyMorphy3 \cite{korobov2015morphological} to convert words into their dictionary forms, reducing data dimensionality while preserving semantic meaning. This approach is particularly effective for the Russian language due to its rich morphology, as it avoids the inaccuracies that stemming introduces by truncating words without context.
        \item \textbf{Stop Word Removal}: excluding overly frequent words that contribute little to the text content using the default stopword list provided by the NLTK package \footnote{https://www.nltk.org} \cite{hardeniya2016natural}, augmented with two Russian pronouns: “which” (\selectlanguage{russian} “который” \selectlanguage{british}) and “such”(\selectlanguage{russian} “такой” \selectlanguage{british}).
    \end{enumerate}
 
\subsection{Neural baseline models}

Neural baseline models used in our work are subdivided into pre-trained dense retrievers (bi-encoders) and rerankers. Bi-encoders generate embeddings for queries and documents and calculate their cosine similarity. Rerankers take a query and a document as an input and calculate the probability of the document to be relevant to the query. Rerankers are applied to the best documents found by lexical or dense retrievers and usually improve the performance of combined retrieval.
Dense retrievers include the following pre-trained bi-encoders:
\begin{itemize}
    \item  LaBSE bi-encoder \cite{feng2022language}. LaBSE was pre-trained with a translation ranking task. This allows to find sentence paraphrases in a single language or different languages.\footnote{https://huggingface.co/sentence-transformers/LaBSE} 
    \item Multilingual E5 in three sizes: large \footnote{https://huggingface.co/intfloat/multilingual-e5-large}, base \footnote{https://huggingface.co/intfloat/multilingual-e5-base} and small \footnote{https://huggingface.co/intfloat/multilingual-e5-small} \cite{wang2024multilingual}. The multilingual E5 model was trained on a large multilingual corpus using a weakly supervised contrastive pretraining method with InfoNCE contrastive loss. Then it was fine-tuned on high-quality labeled multilingual datasets for retrieval tasks. 
   \item BGE-M3 model \footnote{https://huggingface.co/BAAI/bge-m3} \cite{chen2024bge}. The BGE-M3 model was pre-trained on a large multilingual and cross-lingual unsupervised data, and subsequently fine-tuned on multilingual retrieval datasets using a custom loss function based on the InfoNCE loss function.
    \item USER-BGE-M3 \footnote{https://huggingface.co/deepvk/USER-bge-m3}. USER-BGE-M3 is a sentence-transformer model for training embeddings for Russian. The model is initialized from the en-ru-BGE-M3 model \footnote{https://huggingface.co/TatonkaHF/bge-m3\_en\_ru},  a shrinked version of the BGE-M3 model, and then trained on the Russian datasets.
    \item ru-en-RoSBERTa\footnote{https://huggingface.co/ai-forever/ru-en-RoSBERTa} \cite{snegirev2024russian}.
    ruRoBERTa model \cite{zmitrovich2024family} \footnote{https://huggingface.co/ai-forever/ruRoberta-large} was used as a basic model and then RoSBERTa embeddings were fine-tuned on Russian and English datasets. 
\end{itemize}
    
As a reranker, we use the bge-reranker-v2-m3 reranker \footnote{https://huggingface.co/BAAI/bge-reranker-v2-m3}. In our work, we use BGE models with a max-length parameter set to 2048.

\begin{table}[h]
    \centering
    \begin{tabular}{llrrrl}
    \hline
    \textbf{Model} & \textbf{Based on} & \textbf{Parameters} & \textbf{Dim} & \textbf{Max input} \\ 
    \toprule
    Multilingual-E5-large & XLM-RoBERTa-large  & 560M & 1024 & 512 \\ 
    Multilingual-E5-base & XLM-RoBERTa-base & 278M & 768 & 512 \\
    Multilingual-E5-small & Multilingual-MiniLM & 118M & 384 & 512 \\ 
    BGE-M3 & BGE-M3 & 568M & 1024 & 8192\\
    USER-BGE-M3 & BGE-M3 & 359M & 1024 & 8192 \\
    RoSBERTa & SBERT & 404M & 1024 & 512 \\
    LaBSE & LaBSE & 471M & 768 & 256\\
    \midrule
    bge-reranker-v2-m3 & BGE-M3 & 568M & 1024 & 8192 \\ 
    \bottomrule
    \end{tabular}
    \caption{Model Specifications and Details}
    \label{tab:model_specs}
\end{table}

 \section{Results}

We evaluated the models on the RusBEIR datasets using NDCG@10, MAP@10, and Recall@10. Since all metrics showed similar trends, we present only the NDCG@10 results in the table below for brevity. Additional details on MAP@10 and Recall@10 are available in the Additional Metrics section.

\begin{table}[ht]
    \centering
    \resizebox{\linewidth}{!}{
    \begin{tabular}{lc|ccccccc|ccc}
        \toprule
        Model (→) & \multicolumn{1}{c}{Lexical} & \multicolumn{7}{c}{Dense} & \multicolumn{3}{c}{Re-ranking} \\
        \cmidrule(lr){1-1} \cmidrule(lr){2-2} \cmidrule(lr){3-9} \cmidrule(lr){10-12}
        Dataset (↓) & BM25 & mE5-large & mE5-base & mE5-small & BGE-M3 & USER-BGE-M3 &  RoSBERTa & LaBSE & BM25+BGE & mE5-large+BGE & BGE-M3+BGE \\
        \midrule
        rus-NFCorpus & \underline{32.33} & 30.96 & 26.90 & 26.79 & 30.86 & 30.28 & 27.24 & 18.53 & \textbf{34.83} & 33.18 & 32.46 \\ 
        rus-ArguAna & 41.49 & 49.06 & 39.40 & 39.59 & \underline{50.75} & 46.52 & 49.38 & 25.52 & 52.91 & \textbf{54.01} & 53.87 \\     
        rus-SciFact & \underline{65.60} & 63.49 & 63.46 & 60.46 & 62.42 & 58.25 & 53.90 & 29.07 & 70.40 & \textbf{71.34} & 69.64\\
        rus-SCIDOCS &  13.99 & 13.47 & 12.09 & 10.60 & \underline{15.04} & 14.46 & 14.43 & 8.17 & 15.31 & 15.98 & \textbf{16.21} \\
        \midrule
        rus-MMARCO & 15.25 & \underline{34.04} & 30.27 & 29.07 & 29.51 & 27.92 & 20.16 & 9.06 & 24.12 & \textbf{36.95} & 34.52 \\
        rus-MIRACL & 25.13 & 66.99 & 61.41 & 58.52 & \underline{70.50} & 67.23 & 53.11 & 15.70 & 41.51 & 75.90 & \textbf{76.44} \\
        rus-XQuAD & 96.19 & \underline{97.33} & 95.84 & 95.66 & 95.97 & 95.63 & 93.90 & 69.77 & 98.85 & \textbf{98.97} & \textbf{98.97} \\
        rus-XQuAD-Sentences & 82.36 & \underline{88.84} & 86.37 & 85.41 & 86.91 & 85.42 & 83.20 & 75.33 & 89.93 & \textbf{92.08} & 91.69\\
        rus-TyDi QA & 35.80 & \underline{59.41} & 55.91 & 55.23 & 58.34 & 57.86 & 52.06 & 28.05 & 50.12 & \textbf{66.20} & 65.78\\
        \midrule
        SberQuad-retrieval & 68.19 & 67.11 & 65.13 & 61.03 & \underline{68.26} & 67.03 & 63.59 & 37.54 & \textbf{70.34} & 69.41 & 68.21 \\
        ruSciBench-retrieval & 36.69 & 50.81 & 45.74 & 42.93 & \underline{55.85} & 53.58 & 44.89 & 17.93 & 49.93 & 65.33 & \textbf{69.05}  \\
        ru-facts & 92.56 & 93.65 & 93.55 & 93.06 & \textbf{93.91} & 93.77 & 93.66 & 93.10 & 92.72 & 92.87 & 92.87 \\
        RuBQ & 37.33 & \underline{74.11} & 69.63 & 68.60 & 71.26 & 70.00 & 66.81 & 30.59 & 56.90 & \textbf{77.03} & 76.00 \\
        Ria-News & 64.63 & 80.67 & 70.24 & 70.00 & 82.99 & \underline{83.52} & 78.85 & 61.57 & 78.12 & 86.22 & \textbf{86.85} \\
        \midrule
        wikifacts-articles & \underline{84.28} & 66.09 & 63.04 & 67.86 & 74.50 & 79.41 & 74.13 & 45.17 & \textbf{85.25} & 83.06 & 83.91 \\
        wikifacts-para & \underline{61.31} & 50.15 & 49.51 & 34.71 & 54.55 & 57.53 & 50.66 & 14.78 & \textbf{66.61} & 59.95 & 63.76\\
        wikifacts-sents & 33.64 & 35.90 & 30.75 & 22.57 & 37.59 & 34.90 & \textbf{40.59} & 25.79 & 39.96 & 38.53 & 39.20 \\
        \midrule
        \textbf{Avg} & 52.16 & 60.12 & 56.43 & 54.24 & \underline{61.13} & 60.19 & 56.50 & 35.63 & 59.87 & 65.71 & \textbf{65.85} \\
        \bottomrule
    \end{tabular}
    }
    \caption{Performance comparison across different models and datasets. The best results for each dataset are in bold; the results of the best single models are underlined.}
    \label{tab:ndcg10-results} 
\end{table}

The analysis of Table \ref{tab:ndcg10-results} indicates that the best performance on the benchmark is achieved through the combination of the BGE-M3 model and the BGE reranker. Notably, the combination of mE5-large bi-encoder with  the BGE reranker yields close results. Among the individual models, the mE5-large bi-encoder and both multilingual BGE variants stand out as top performers, surpassing BM25 by an average margin of 15.9 percentage points.


Overall, LaBSE performs the worst among all the models presented. This can be attributed to its training objective, which focuses on finding similar sentences across different languages or paraphrases within the same language. As a result, when confronted with queries that lack lexical overlap with sentences in the corpus, its performance drops.

RoSBERTa model performs on par with mE5-base and mE5-small, but the size of mE5-base (278M) againts RoSBERTa (404M) makes mE5-base more preferable to use.

At the same time, it is worth noting that the BM25 model is the best single model on four datasets: rus-NFCorpus, rus-SciFact, wikifacts-articles and  wikifacts-para. 
The best results on these datasets, as well as others where single BM25 performed only slightly worse than neural retrievers, are achieved by combining BM25 with the BGE reranker. Three of the datasets with a significant BM25 margin contain longer documents than the average in the benchmark. On the wikifacts-articles dataset, which includes full-text documents, BM25 outperforms the BGE-M3 model by 13 percentage points and the mE5-large model by 27 percentage points. This highlights a limitation of the mE5 models in retrieving long documents due to their small maximum input size (512 tokens). Additionally, the rus-NFCorpus and rus-SciFact datasets are domain-specific, which may result in lower-quality multilingual vector representations compared to general datasets.

Furthermore, it should be noted that the results of the BGE models presented in Table \ref{tab:ndcg10-results} were obtained with a maximum input length set to 2048. However, as indicated in Table \ref{tab:model_specs}, BGE models can process up to 8192 tokens, making them more suitable for full-text search in long documents.

Our experiments demonstrated that BM25 remains a strong baseline for information retrieval, particularly for full-document retrieval. Neural models, especially mE5-large and BGE-M3, achieved the best results on the benchmark and confirmed the findings of other BEIR-based studies \cite{thakur2021beir,wang2022text}.

\section{Conclusion}
In this paper, we introduced RusBEIR, a comprehensive BEIR-inspired benchmark designed for the zero-shot evaluation of information retrieval (IR) models in the Russian language. Consisting of 17 datasets from diverse domains and tasks, RusBEIR integrates adapted datasets from existing benchmarks alongside novel datasets to further enrich its collection. By providing a large-scale resource compatible with the original BEIR format, RusBEIR enables systematic evaluation and comparison of both lexical and neural IR models, with a particular emphasis on zero-shot performance.

Our study stresses the importance of accurate preprocessing, particularly for lexical models, where preprocessing significantly impacts the performance in morphologically rich languages as Russian. Additionally, we introduced a series of Russian Wikipedia-based datasets that further expand the scope of RusBEIR, enabling more granular exploration of IR performance across document lengths and tasks.

The results of our experiments confirm that BM25 remains a robust baseline for full-document retrieval, while state-of-the-art neural models, such as mE5-large and BGE-M3, demonstrate superior performance on most datasets. These findings are consistent with previous BEIR-based studies and underscore the advantages of neural approaches, particularly when unprocessed data are used as input. However, our analysis also highlights certain limitations of neural models, such as challenges with long-document retrieval due to input size constraints. The efficiency comparison between BM25 and neural models such as mE5 and BGE remains an open question and will be explored further in future research.

By providing a unified framework and detailed insights into the comparative performance of lexical and neural models, we hope RusBEIR will serve as a valuable tool for advancing research and innovation in information retrieval for the Russian language.

\section*{Acknowledgements}
The work is supported by the Russian Science Foundation under Agreement No. 25-21-00206. \\ The research was carried out using the MSU-270 supercomputer of Lomonosov Moscow State University.

\printbibliography

\section{Additional metrics}



\subsection{MAP}

Mean Average Precision (MAP) is used to assess the overall precision of a retrieval system across multiple queries. It computes the average precision for each query and then takes the mean across all queries. MAP provides a single summary measure that reflects both the ranking quality and the system's ability to retrieve relevant documents.

The MAP@10 results obtained from the models’ inference on the benchmark datasets are shown below.

\begin{table}[h]
    \centering
    \resizebox{\linewidth}{!}{
    \begin{tabular}{lc|ccccccc|ccc}
        \toprule
        Model (→) & \multicolumn{1}{c}{Lexical} & \multicolumn{7}{c}{Dense} & \multicolumn{3}{c}{Re-ranking} \\
        \cmidrule(lr){1-1} \cmidrule(lr){2-2} \cmidrule(lr){3-9} \cmidrule(lr){10-12}
        Dataset (↓) & BM25 & mE5-large & mE5-base & mE5-small & BGE-M3 & USER-BGE-M3 &  RoSBERTa & LaBSE & BM25+BGE & mE5-large+BGE & BGE+BGE \\
        \midrule
        rus-NFCorpus & \underline{12.52} & 11.39 & 9.23 & 9.37 & 11.40 & 10.99 & 10.02 & 5.74 & \textbf{13.47} & 12.62 & 12.33 \\ 
        rus-ArguAna & 32.76 & 40.71 & 31.69 & 32.08 & \underline{41.85} & 37.57 & 40.32 & 20.40 & 45.32 & \textbf{45.48} & 45.12 \\ 
        rus-SciFact & 61.47 & \underline{59.76} & 58.84 & 55.67 & 57.82 & 53.60 & 49.25 & 25.91 & 67.10 & \textbf{67.72} & 66.25\\
        rus-SCIDOCS & 8.03 & 7.53 & 6.69 & 5.89 & \underline{8.64} & 8.29 & 8.27 & 4.45 & 8.88 & 9.28 & \textbf{9.37} \\
        \midrule
        rus-MMARCO & 11.88 & \underline{28.11} & 24.94 & 23.82 & 24.03 & 22.59 & 16.03 & 07.07 & 21.30 & \textbf{30.78} & 29.13 \\
        rus-MIRACL & 18.61 & 56.64 & 50.90 & 48.00 & \underline{60.52} & 57.11 & 42.36 & 10.94 & 35.79 & 67.24 & \textbf{67.77}\\
        rus-XQuAD & 95.04 & \underline{96.11} & 94.63 & 94.37 & 94.81 & 94.35 & 92.24 & 65.84 & 98.57 & 98.64 & \textbf{98.64}\\
        rus-XQuAD-Sentences & 79.32 & \underline{85.89} & 83.15 & 82.03 & 83.80 & 81.99 & 79.32 & 71.12 & 88.44 & \textbf{90.08} & 89.75 \\
        rus-TyDi QA & 30.16 & \underline{51.78} & 48.79 & 48.35 & 51.02 & 50.90 & 44.88 & 22.91 & 46.13 & \textbf{59.50} & 59.18 \\
        \midrule
        SberQuad-retrieval & 58.36 & 57.43 & 55.95 & 50.94 & \underline{60.25} & 58.81 & 55.38 & 30.49 & \textbf{60.84} & 59.96 & 58.90 \\
        ruSciBench-retrieval & 27.07 & 39.31 & 34.50 & 31.50 & \underline{43.30} & 41.47 & 33.72 & 12.48 & 40.43 & 54.74 & \textbf{58.12} \\
        ru-facts & 90.03 & 91.66 & 91.30 & 90.66 & \underline{91.79} & 91.60 & 91.47 & 90.70 & 90.24 & 90.39 & \textbf{90.39} \\
        RuBQ & 29.36 & \underline{66.24} & 61.94 & 60.95 & 63.84 & 62.29 & 58.72 & 24.30 & 51.64 & \textbf{70.10} & 69.25\\
        Ria-News & 60.41 & 75.94 & 65.67 & 65.59 & 79.94 & \underline{80.62} & 75.44 & 57.79 & 76.75 & 84.17 & \textbf{84.74}\\
        \midrule
        wikifacts-articles & 78.60 & 65.96 & 55.80 & 60.32 & 68.32 & \underline{73.51} & 67.09 & 37.95 & \textbf{80.44} & 78.59 & 79.34 \\
        wikifacts-para & 50.67 & 42.50 & 39.54 & 26.52 & 44.00 & \underline{46.87} & 40.43 & 10.01 & \textbf{56.71} & 51.09 & 54.11 \\
        wikifacts-sents & 24.45 & 29.53 & 22.52 & 16.17 & 27.44 & 25.01 & 29.84 & 18.15 & 30.50 & 28.38 & 28.60 \\
        \midrule
        \textbf{Avg} & 45.22 & 53.32 & 49.18 & 47.19 & \underline{53.69} & 52.80 & 49.10 & 30.37 & 53.68 & 58.75 & \textbf{58.88} \\
        \bottomrule
    \end{tabular}
    }
    \caption{Performance comparison across different models and datasets. The best results for each dataset are in bold; the results of the best single models are underlined.}
    \label{tab:map10-results} 
\end{table}

\subsection{Recall}

Recall quantifies the proportion of relevant documents that are successfully retrieved by the system. It is defined as the ratio of the number of relevant documents retrieved to the total number of relevant documents available. In the context of information retrieval, high recall is crucial to ensure that the system does not miss important information.

The Recall@10 results obtained from the models’ inference on the benchmark datasets are shown below.

\begin{table}[ht]
    \centering
    \resizebox{\linewidth}{!}{
    \begin{tabular}{lc|ccccccc|ccc}
        \toprule
        Model (→) & \multicolumn{1}{c}{Lexical} & \multicolumn{7}{c}{Dense} & \multicolumn{3}{c}{Re-ranking} \\
        \cmidrule(lr){1-1} \cmidrule(lr){2-2} \cmidrule(lr){3-9} \cmidrule(lr){10-12}
        Dataset (↓) & BM25 & mE5-large & mE5-base & mE5-small & BGE-M3 & USER-BGE-M3 &  RoSBERTa & LaBSE & BM25+BGE & mE5-large+BGE & BGE+BGE \\
        \midrule
        rus-NFCorpus & \underline{16.09} & 15.68 & 12.56 & 12.79 & 14.93 & 14.56 & 13.17 & 8.57 & \textbf{16.69} & 15.59 & 14.97 \\ 
        rus-ArguAna & 69.70 & 75.82 & 64.30 & 63.87 & \underline{79.16} & 75.32 & 78.52 & 42.11 & 76.81 & 81.01 & \textbf{81.65}\\ 
        rus-SciFact & 76.63 & \underline{76.88} & 76.42 & 73.46 & 75.08 & 70.90 & 66.61 & 37.71 & 79.39 & \textbf{80.88} & 78.58\\
        rus-SCIDOCS & 14.48 & 14.14 & 12.80 & 11.14 & \underline{15.59} & 14.88 & 15.34 & 8.33 & 15.66 & 16.34 & \textbf{17.02} \\
        \midrule
        rus-MMARCO & 25.77 & \underline{52.38} & 46.68 & 45.36 & 46.53 & 44.42 & 33.02 & 15.26 & 32.32 & \textbf{55.90} & 50.90 \\
        rus-MIRACL & 31.32 & 76.70 & 71.03 & 68.43 & \underline{79.59} & 76.44 & 63.69 & 21.16 & 39.28 & 81.81 & \textbf{82.59} \\
        rus-XQuAD & 99.58 & \underline{99.75} & 99.50 & 99.50 & 99.41 & 99.41 & 98.91 & 82.02 & 99.66 & \textbf{99.92} & \textbf{99.92} \\
        rus-XQuAD-Sentences & 91.78 & \underline{97.44} & 96.09 & 95.76 & 96.30 & 95.88 & 95.13 & 88.40 & 94.31 & \textbf{98.07} & 97.48 \\
        rus-TyDi QA & 51.26 & \underline{79.03} & 75.34 & 73.94 & 78.31 & 76.55 & 71.56 & 42.43 & 59.07 & \textbf{83.88} & 82.80 \\
        \midrule
        SberQuad-retrieval & \underline{96.47} & 93.47 & 92.14 & 90.71 & 91.94 & 91.42 & 88.15 & 58.84 & \textbf{97.32} & 96.29 & 94.70 \\
        ruSciBench-retrieval & 38.63 & 53.98 & 47.15 & 45.40 & \underline{57.87} & 54.68 & 46.91 & 19.23 & 45.64 & 62.51 & \textbf{66.62}\\
        ru-facts & 99.82 & \textbf{100.00} & \textbf{100.00} & 99.96 & \textbf{100.00} & \textbf{100.00} & 99.96 & \textbf{100.00} & 99.82 & \textbf{100.00} & \textbf{100.00}\\
        RuBQ & 52.71 & \underline{86.26} & 83.10 & 81.20 & 84.06 & 83.18 & 81.29 & 41.47 & 62.00 & \textbf{88.84} & 86.90 \\
        Ria-News & 77.84 & 90.30 & 84.50 & 83.73 & 92.34 & \underline{92.42} & 89.41 & 73.41 & 82.19 & 92.44 & \textbf{93.24}\\
        \midrule
        wikifacts-articles & \underline{88.39} & 77.49 & 71.44 & 77.90 & 82.39 & 86.72 & 82.99 & 53.21 & \textbf{90.64} & 85.84 & 88.98\\
        wikifacts-para & \underline{67.30} & 60.51 & 57.65 & 41.44 & 62.87 & 65.81 & 59.37 & 17.87 & \textbf{70.17} & 62.48 & 67.77\\
        wikifacts-sents & 35.42 & 40.65 & 32.28 & 24.63 & 39.89 & 37.39 & \underline{\textbf{43.22}} & 26.85 & 39.83 & 40.10 & 41.54\\
        \midrule
        \textbf{Avg} & 60.78 & 70.03 & 66.06 & 64.07 & \underline{70.37} & 69.41 & 66.31 & 43.35 & 64.75 & 73.05 & \textbf{73.27} \\
        \bottomrule
    \end{tabular}
    }
    \caption{Performance comparison across different models and datasets. The best results for each dataset are in bold; the results of the best single models are underlined.}
    \label{tab:recall10-results} 
\end{table}




\end{document}